\documentclass[10pt]{iopart}
\usepackage{epsfig}
\begin{document}

\title{Review of SIS Experimental Results on Strangeness}

\author{Helmut Oeschler}

\address{Institut f\"ur Kernphysik, Technische Universit\"at Darmstadt, \\
D-64289 Darmstadt, Germany}


\begin{abstract}
A review of meson emission in heavy ion collisions at incident energies 
around 1 -- 2 $A\cdot$GeV is presented. It is shown how
the shape of the spectra and the various particle yields
vary with system size, with centrality and with incident energy.
A statistical model assuming thermal and chemical equilibrium and exact
strangeness conservation (i.e.~strangeness conservation per collision)
explains most of the observed features.

Emphasis is put onto the study of $K^+$ and $K^-$ emission.
In the framework of this statistical model it is shown that the experimentally
observed equality of $K^+$ and $K^-$ rates at threshold corrected energies 
$\sqrt{s} - \sqrt{s_{th}}$ is due to a crossing of two excitation functions.
Furthermore, the independence of the $K^+$ to $K^-$ ratio on the number of
participating nucleons observed between 1 and 10 $A\cdot$GeV is consistent with
this model.
The observed flow effects are beyond the scope of this model.
\end{abstract}

\section{Introduction}
Central heavy ion collisions
at relativistic incident energies represent an ideal tool to study
nuclear matter at high densities and at high temperatures.
Particle production is at all incident energies a key quantity to extract
information on the properties of nuclear matter under theses extreme conditions.
Particles carrying strangeness have turned out to be very valuable messengers.
At incident energies around 1 -- 2 $A\cdot$GeV, as available at the SIS 
accellerator at GSI, Darmstadt, these particles are produced below their
threshold in $NN$ collisions. Due to strangeness conservation, a  $K^+$ 
meson is produced 
with an associated $\Lambda$, having a threshold of 671 MeV in the 
centre-of-mass system
and 1.58 GeV in the laboratory system. $K^-$ is ``cheapestly'' produced 
by $K^- K^+$ pair creation, 
yielding thresholds of 987 MeV (c.m.~system) and 2.5 GeV (lab system).
The subthreshold strangeness production is the key parameter 
governing this paper.
Furthermore, $K^-$ and $K^+$ behave very differently in nuclear matter.
While $K^+$ hardly finds a partner to react with, $K^-$ exhibits a strong
interaction with nuclear matter.

In this paper, an 
overview of the shape of the measured meson spectra is given and their
yields are studied as a function of system size $A+A$, of the 
number of participating nucleons $A_{part}$ and as a function of incident 
energy.
These results are discussed along with the question whether the observations
are in agreement with the assumption of a thermal and chemical equilibrium.
Of special interest is the ratio of $K^+/K^-$ yields as the measured values
in heavy ion collisions differ strongly from those in pp reactions.
These findings have lead to the interpretation that in heavy ion collisions
the ``effective masses'' of $K^+$ and $K^-$ might be changed as predicted
for dense nuclear matter.
It turns out that a statistical approach is very successful to explain the 
measured yields.

\section{Experimental Devices}

At GSI, Darmstadt, three major experimental devices for
measuring strange particle production are installed:
\begin{itemize}
\item The {\bf FOPI} setup, a
$4\pi$ detector, designed to study the global features of heavy ion collisions
at these energies \cite{fopi}. This detector measures all particles
and is optimized for flow studies and informations related with
the reaction plane. It is also suited for vertex reconstruction 
to detect $K^0$, $\Lambda$, etc.
In the present setup kaons can only be detected 
at backward rapidities.
A major upgrade is under way to improve the detection and trigger capabilities.
\item The magnetic Kaon Spectrometer {\bf KaoS} specially designed to study
rare kaon production \cite{kaos}. 
It is optimized for short flight paths because of the kaon decay.
It has a very selective trigger to handle high beam intensities.
Multiwire chambers for tracking are installed to reduce
the background in kaon measurements with  $K^+$ rates 
$10^5$ times smaller than for protons. 
Two detector arrays serve for centrality selection and 
reaction-plane determination.
Most measurements are performed around 
midrapidity; in some cases a large fraction of the 
full phase space has been studied 
by moving the spectrometer.
\item The {\bf TAPS} detector consists of six arrays of BaF$_2$ modules 
measuring $\gamma$ rays. It detects among others the $\eta$ which decays
by $\gamma$ emission. $\eta$ has net strangeness zero, but contains 
$s \bar s$ pairs.  
\end{itemize}
                      
Since this experimental review is restricted to spectral shapes and yields,
mainly results from the KaoS collaboration are shown. Very interesting
flow studies
are given in the two following talks by P.~Crochet and by Y.~Shin.

\section{Spectra and Yields}

\begin{figure}[h]
\begin{minipage}[t]{5.3cm}
\mbox{\epsfig{width=9cm,file=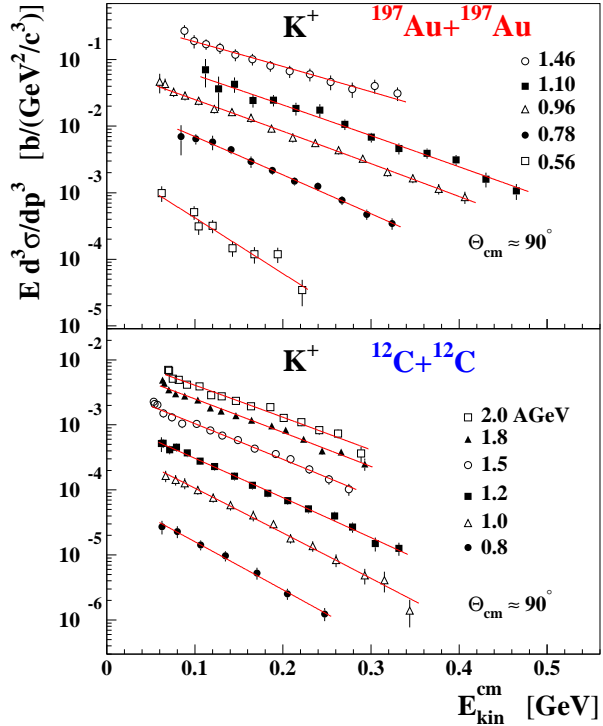}}
\end{minipage}
\hfill
\begin{minipage}[b]{5.6cm}
\vspace*{-3cm}
\caption{$K^+$ spectra measured at midrapidity for Au+Au (upper part)
and C+C (lower part) at various incident energies~\cite{Laue, CS}.
The solid lines are Boltzmann fits to the data.}
\label{Kplus}
\end{minipage}
\end{figure}

Spectra of $K^+$ from mass symmetric systems C+C and Au+Au
at incident energies from 0.6 to 2.0 $A\cdot$GeV 
measured at midrapidity are shown in Fig.~\ref{Kplus},
exhibiting Boltzmann shapes~\cite{Laue, CS}.
Their inverse slope parameters increase 
monotoneously with incident energy
and the heavier system exhibits harder spectra than the light system at the 
same incident energy.

Pion spectra show deviations from a Boltzmann shape~\cite{MUE97, 
Pelte97, HO_MAZ}.
Their  shapes are qualitatively understood as composed of two parts,
a direct (or thermal) pion contribution and pions from decaying baryonic
resonances, which produce an enhancement at lower kinetic 
energies~\cite{WEI98, Hong, FOE98}. The slopes of the high-energy component 
agree well with those extracted from the corresponding $K^+$ spectra.

\begin{figure}[h]
\begin{center}
\mbox{\epsfig{width=6.4cm,file=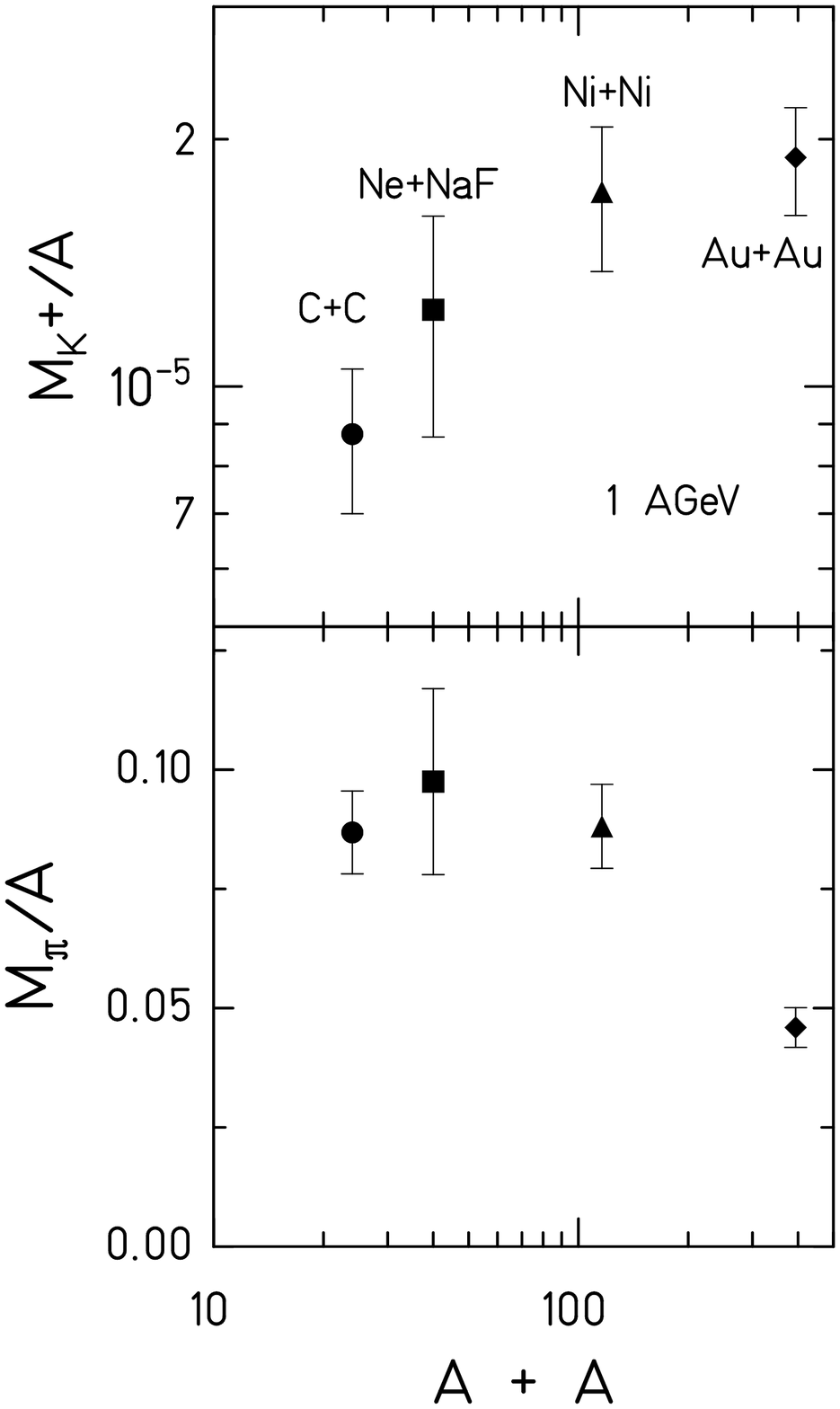}}
\mbox{\epsfig{width=6.4cm,file=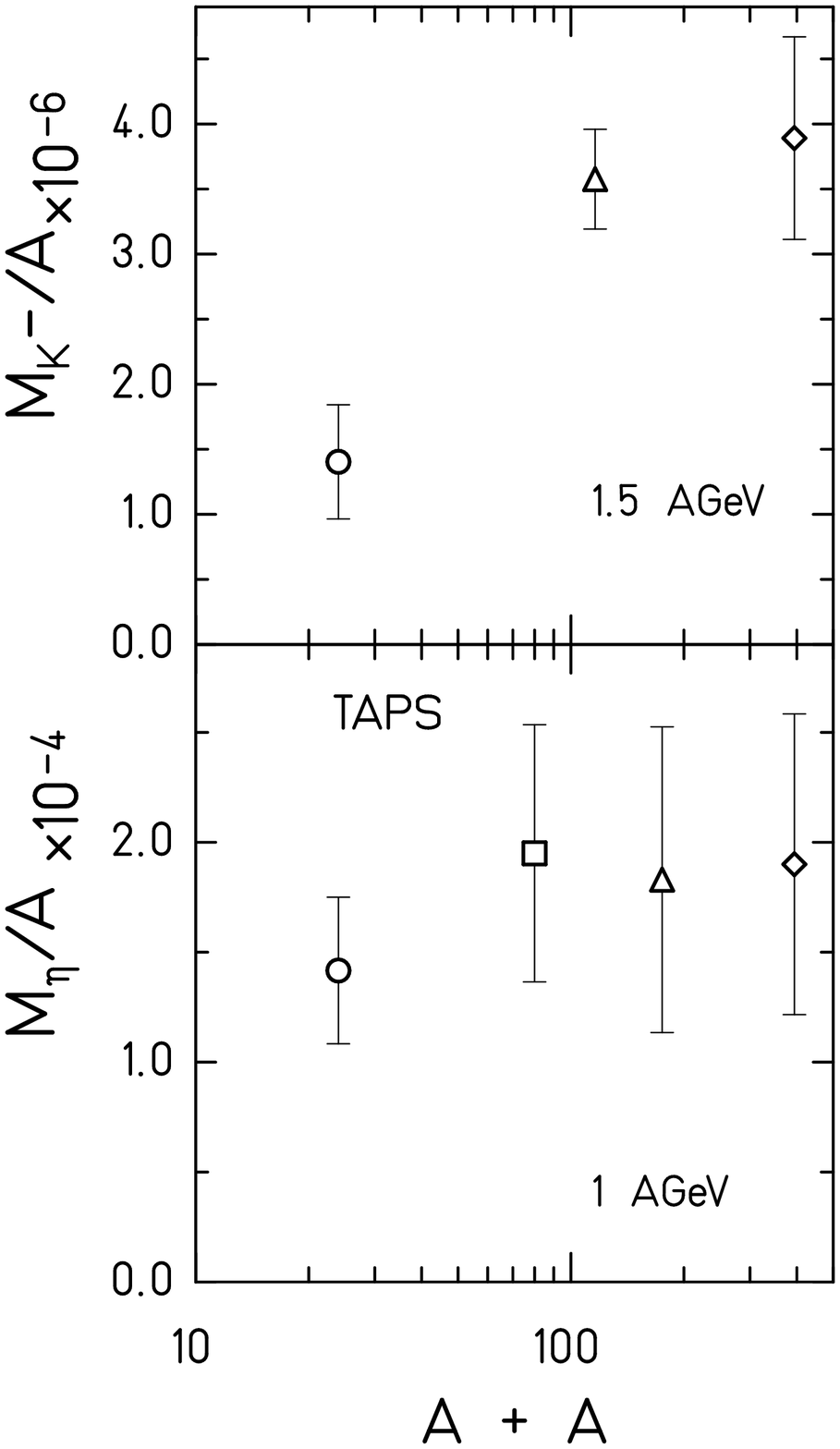}}
\end{center}
\caption{Multiplicities per mass number $A$ as a function of $A+A$
for $K^+$ and 
pions ($\pi^+, \pi^0, \pi^-$) (at 1 $A\cdot$GeV) on the r.h.s 
and $K^-$ (at 1.5 $A\cdot$GeV) and $\eta$ (at 1 $A\cdot$GeV) left.
Partially preliminary data from~\cite{Laue, CS, MUE97, Barth, AF, eta}.
Particles with strangeness exhibits a strong rise while those with s=0 do not.}
\label{K_PI_AA}
\end{figure}

Next, we study the inclusive yields. In many cases the phase-space distribution
has been measured, yielding slight anisotropies. These effects are taken 
into account
to obtain the inclusive cross sections and are in the order of at most 30\%
as compared to a mid-rapidity measurement extrapolated to 4$\pi$ 
assuming isotropic emission~\cite{Laue,CS}.
These cross sections are divided by the geometrical reaction cross sections
(using $r_0$=1.2 fm)
to obtain multiplicities. For a comparison of different systems
these multiplicity values are divided by the system mass $A$ as shown 
in Fig.~\ref{K_PI_AA}
for 1  and 1.5 $A\cdot$GeV.
A striking difference is seen between the trends observed for $K^+$ and $K^-$
which exhibit a strong rise of $M/A$ with system size and
pions which show a slight decrease.    The values observed for $\eta$
production seem to be rather constant. Some $\eta$ measurements were not fully
inclusive. Those values have been corrected.
Having in mind that kaons are produced below their $NN$ thresholds,
one can think of mechanisms causing such a rise: 
Only multiple collisions can collect the energy needed for kaon production and
the probability for multistep interactions increases 
with density and/or system size.
  
\begin{figure}[h]
\begin{minipage}[t]{5.3cm}
\mbox{\epsfig{width=7.7cm,file=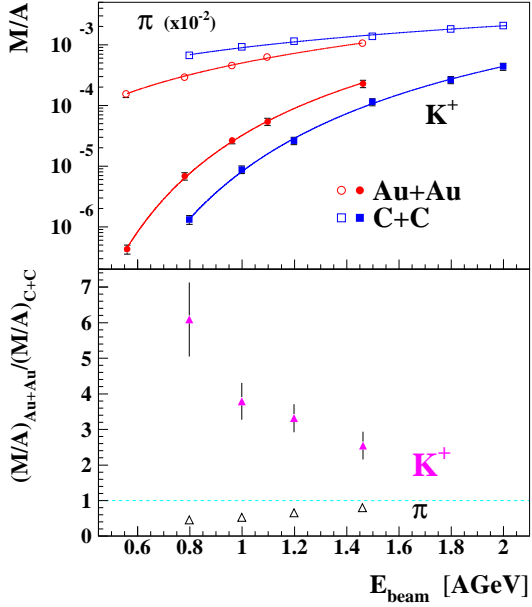}}
\vspace*{-3cm}
\end{minipage}
\hfill
\begin{minipage}[b]{6.6cm}
\caption{Upper part: Multiplicities of $K^+$ and pions per $A$ 
as a function of incident energy. 
The pion data include charged and neutral pions.
The lines represent fits to the data (see~\cite{CS}).
Lower part: Ratio of the multiplicities per $A$ (Au+Au over C+C) as a function
of incident energy.}
\label{EOS}
\end{minipage}
\end{figure}

It is of interest to follow how the mass dependence varies with incident energy.
The upper part of Fig.~\ref{EOS} 
shows the pion and $K^+$ multiplicities per $A$ for C+C and Au+Au
collisions as a function of incident energy. 
The pion multiplicities ($M_{\pi}/A$) representing
the sum of charged and neutral pions~\cite{CS}, 
are smaller for Au+Au than for C+C
whereas the $K^+$ multiplicities per $A$  exhibit the opposite behaviour.
These trends are already seen in Fig.~\ref{K_PI_AA}. With increasing incident 
energy the differences in $M/A$ between the light and the heavy system 
become smaller as can be seen in Fig.~\ref{EOS}.

The mass dependence of the $K^+$ yields is
demonstrated in the lower
part of Fig.~\ref{EOS} showing the ratio $(M/A)_{\rm Au+Au}/(M/A)_{\rm C+C}$.
The enhancement factor for $K^+$ reaches 6 at the lowest measured 
incident energy.
At lower incident energies more and more individual $NN$ interactions
are needed to accumulate the energy to produce a $K^+$ and hence a increasing
sensitivity to density occurs. These data represent therefore the ideal 
set of information 
to extract the nuclear equation of state (EOS). 
Depending on the stiffness of the EOS different densities might be reached.
A soft EOS allows to compress the nuclear matter easier, 
higher densities can be 
achieved and as a consequence more $K^+$ can be produced. 
Yet, qualitative
arguments are not sufficient to extract the stiffness parameter; transport model
calculations are needed and they show that these data are in agreement 
with a soft ($k$= 200 MeV) EOS~\cite{Fuchs}. This conclusion, however,
is put in question by a recent study~\cite{Aichelin}.  

\begin{figure}[h]
\begin{minipage}[t]{5.1cm}
\mbox{\epsfig{width=8.0cm,file=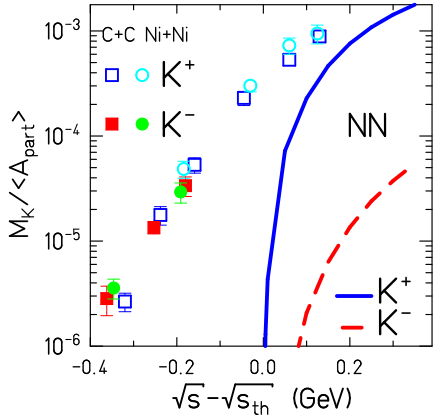}}
\vspace*{-3cm}
\end{minipage}
\hfill
\begin{minipage}[b]{6.9cm}
\vspace*{-3cm}
\caption{Measured $K^+$ and $K^-$ yields in heavy ion and $NN$ collisions
as a function of  $\sqrt{s}-\sqrt{s_{th}}$~\protect{\cite{Laue,Barth,Marc}}.
$<A_{part}>$ is $A/2$ for heavy ion data and 2 for $NN$ collisions.}
\label{KP_KM_sthr}
\end{minipage}
\end{figure}

Negatively charged kaons are at SIS energies always produced 
below their $NN$ threshold
and rather small cross sections are expected in heavy ion collisions.
Furthermore, $K^-$ can be easily 
absorbed in nuclear matter. Yet, the measured yields turned out to be 
rather high compared to the yields measured in pp collisions.
These measurements by the KaoS Collaboration
have attracted considerable interest as in heavy ion collisions
the $K^-$ yield compared to the $K^+$ cross section
is much higher than expected from $NN$ collisions \cite{Laue, Barth}.
This is especially evident, if the kaon multiplicities are
plotted as a function of $\sqrt{s} - \sqrt{s_{th}}$ where 
$\sqrt{s_{th}}$ is the energy needed to produced the respective
particle in $NN$ collisions 
taking into account the mass of the associately produced partner.
The obvious contrast between $NN$ and $AA$ collisions,
shown in Fig.~\ref{KP_KM_sthr},
has lead to the interpretation of the results by
in-medium properties which cause e.g.~a lower threshold for
$K^-$ production when produced in dense matter~\cite{Cassing}. 
Of course, the difference between $NN$ and $AA$ collisions as shown 
in Fig.~\ref{KP_KM_sthr}, is not sufficient to conclude on properties of 
kaons in matter. In heavy ion collisions, kaons can be produced by other
channels, e.g.~$\pi \Lambda \rightarrow K^- N$ which are not available in $NN$
collisions. Only by using detailed transport model calculations  
one can conclude on new properties of kaons in matter~\cite{Cassing}.
They demonstrate that
by including secondary collisions and the newly opened channels
the measured $K^-$ yields are not reproduced when using bare kaon 
masses.

\section{Interpretation within a Statistical Model}

The spectral shapes of particles presented so far, are in agreement with the
interpretation within a statistical concept.
The measured multiplicities for pions and $K^+$ (Fig.~\ref{K_PI_AA})
evidence a strong contrast; while the pion multiplicity 
decreases with the mass of the colliding system, the $K^+$ multiplicity rises
strongly. The latter observation seems to be in conflict with a thermal
interpretation, which -- in a naive view -- should give multiplicities per 
mass number $A$ being constant.

\begin{figure}[h]
\begin{minipage}[t]{5.3cm}
\mbox{\epsfig{width=8.0cm,file=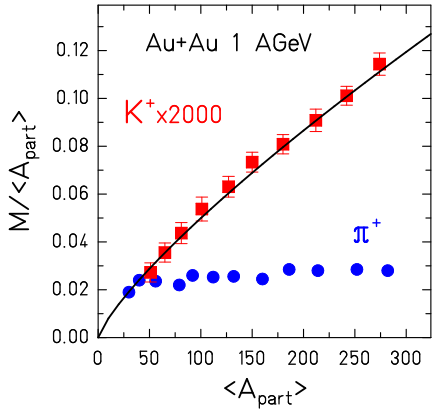}}
\end{minipage}
\hfill
\begin{minipage}[b]{6.9cm}
\vspace*{-3.5cm}
\caption{The multiplicity of $K^+/A_{part}$ rises strongly
with $A_{part}$ in contrast to the pion multiplicity~\protect\cite{Mang}.
This rise can be descibed by the thermal model
including exact strangeness conservation (see text).}
\label{Apart}
\end{minipage}
\end{figure}

Usually, the particle number densities or 
the multiplicities per $A_{part}$, here for pions, are described 
in a simplified way by a Boltzmann factor 
\begin{equation}
\frac{M_{\pi}}{A_{part}}\sim \exp \left(-\frac{<E_{\pi}>}{T}\right).
\end{equation}
The production of strange particles has to fulfil strangeness
conservation. The attempt to describe the measured particle ratios
including strange hadrons at AGS and SPS using a strangeness chemical potential
$\mu_S$ is quite successful~\cite{PBM,Cley}.
However, this grand-canonical treatement is not sufficient,
if the number of produced strange particles is small. 
There, a statistical model has to take care of {\it exact
strangeness conservation} in each reaction as introduced by 
Hagedorn~\cite{Hagedorn}. 
This is done by taking into account
that together with each $K^+$ another strange particle,
e.g.~a $\Lambda$ is produced:
\begin{equation}
\frac{M_{K^+}}{A_{part}}\sim \exp \left(-\frac{<E_{K^+}>}{T}\right)
\left[g_{\Lambda}V \int {d^3p\over (2\pi)^3}
\exp\left(-{{(E_{\Lambda}-\mu_B)}\over T}\right)\right],    
\end{equation} 
with the temperature $T$, the baryo-chemical potential $\mu_B$,
the degeneracy factors $g_i$, the production volume for making the
associate pair $V$ (see~\cite{CLE99, CLE00}) and the total energies $E_i$.
This formula, simplified for demonstration purpose,
neglects other combinations leading to the production
of $K^+$ as well as the use of Bose-Fermi distributions, which are all
included in the computation.
The corresponding formula for $K^-$ production is similar, but does not 
depend on $\mu_B$. This point will become important later on.
\begin{equation}
\frac{M_{K^-}}{A_{part}}\sim \exp \left(-\frac{<E_{K^-}>}{T}\right)
\left[g_{K^+}V \int {d^3p\over (2\pi)^3}
\exp\left(-{E_{K^+}\over T}\right)\right].
\end{equation}
These formulae lead to 
a reduction of $K^+$ and $K^-$ yields as compared to the numbers calculated
without exact strangeness conservation~\cite{CLE99, CLE00}.
Two extreme conditions can be seen from Eqs.~(2) and (3)~\cite{CLE99}.
In the low-number
limit, the additional term (due to the parameter $V$)
leads to a linear rise of $M_{K^+}/A_{part}$ 
while $M_{\pi}/A_{part}$ remains constant.
This is in remarkable agreement with the experimental
observations shown in Fig.~\ref{Apart}. 
For very high temperatures or very large $V$, the terms in brackets
approach unity (see Ref.~\cite{CLE99}) and the formulae coincide
with the grand-canonical procedure.
                  
\begin{figure}
\begin{center}
\vspace*{-1cm}
\hspace*{-.5cm}
\mbox{\epsfig{width=15.1cm,file=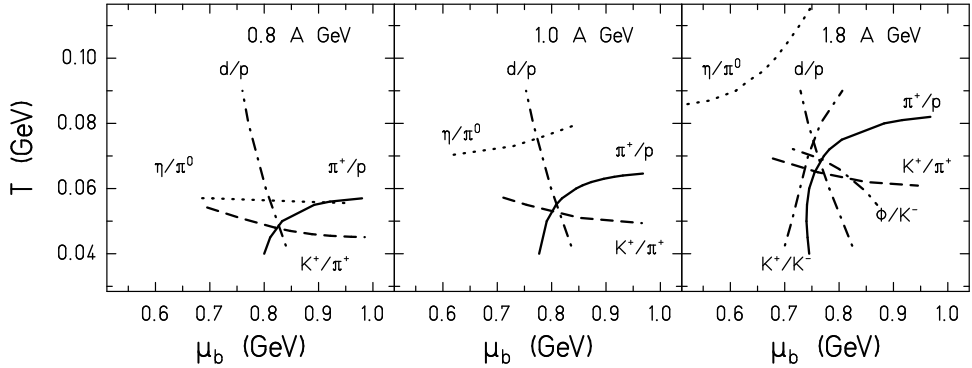}}
\caption{
$T$ versus $\mu_B$ for central Ni+Ni collisions from 0.8~$A\cdot$GeV to
1.8~$A\cdot$GeV. From \protect\cite{CLE99} and new, preliminary data for
$K^+$ and $K^-$ \protect\cite{Marc}.}
\vspace*{-0.7cm}
\label{Therm_ni}
\end{center}
\end{figure}

A further test is the understanding of all particle yields.
The measured yields (or particle ratios)
can be described in this statistical concept by 
combinations of $T$ and $\mu_B$ shown as lines
in Fig.~\ref{Therm_ni} for central Ni+Ni collisions at three incident energies.
Besides the results for $\eta/\pi_0$ all measured particle ratios intersect
within a small area reflecting $T$ and $\mu_B$ at freeze out.
Surprisingly, even the measured $K^+/K^-$ ratio
fits into this representation and this ratio does not depend on the choice
of the volume term $V$. 

\begin{figure}[h]
\begin{minipage}[t]{5.1cm}
\mbox{\epsfig{width=9.4cm,file=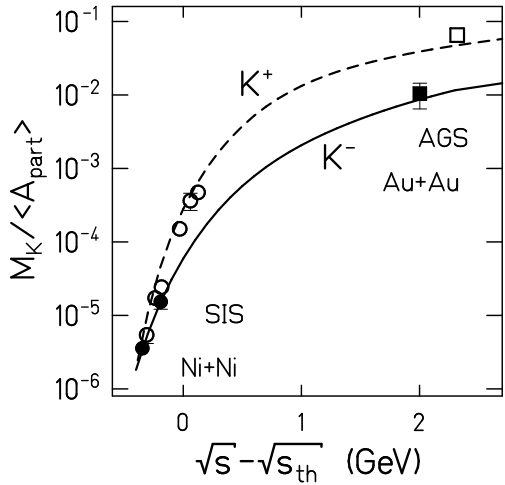}}
\end{minipage}
\hfill
\begin{minipage}[b]{5.9cm}
\vspace*{-3cm}
\caption{{\small Calculated $K^+/A_{part}$ and $K^-/A_{part}$ ratios 
in the thermal model as a function of $\sqrt{s} - \sqrt{s_{th}}$
for Ni+Ni collisions.
The points are results for Ni+Ni collisions at SIS energies 
\protect\cite{Barth, Marc}
and Au+Au
at 10.2 AGeV (AGS)~\protect\cite{Ahle}. At AGS energies the influence of the 
system mass is negligible.}}
\label{KP_KM_sthr_therm}
\end{minipage}
\end{figure}

As the $K^+/K^-$ ratios are in rather good agreement with the assumption of a 
chemical equilibrium, it is of interest to see how the result of the thermal 
concept, using Ref.~\cite{CLEY98},
appears in the representation as a function of $\sqrt{s} - \sqrt{s_{th}}$
(as used in Fig.~\ref{KP_KM_sthr}), which is 
shown in Fig.~\ref{KP_KM_sthr_therm}. 
Note that the x-axis extends now over a much larger range 
than in Fig.~\ref{KP_KM_sthr}.
\vspace*{-1cm}

At values of $\sqrt{s} - \sqrt{s_{th}}$ less than zero,
the two excitation function cross. They 
differ at AGS energies by a factor of five.
The difference in the rise of the two excitation functions 
can be understood by the formulae given above.
The one for $K^+$ production contains ($E_{\Lambda}-\mu_B$) while the other
has $E_{K^+}$ in the exponent. As these two values are different, the
excitation functions, i.e.~the variation with $T$, exhibit different rises.

\begin{figure}
\mbox{\epsfig{width=7.2cm,file=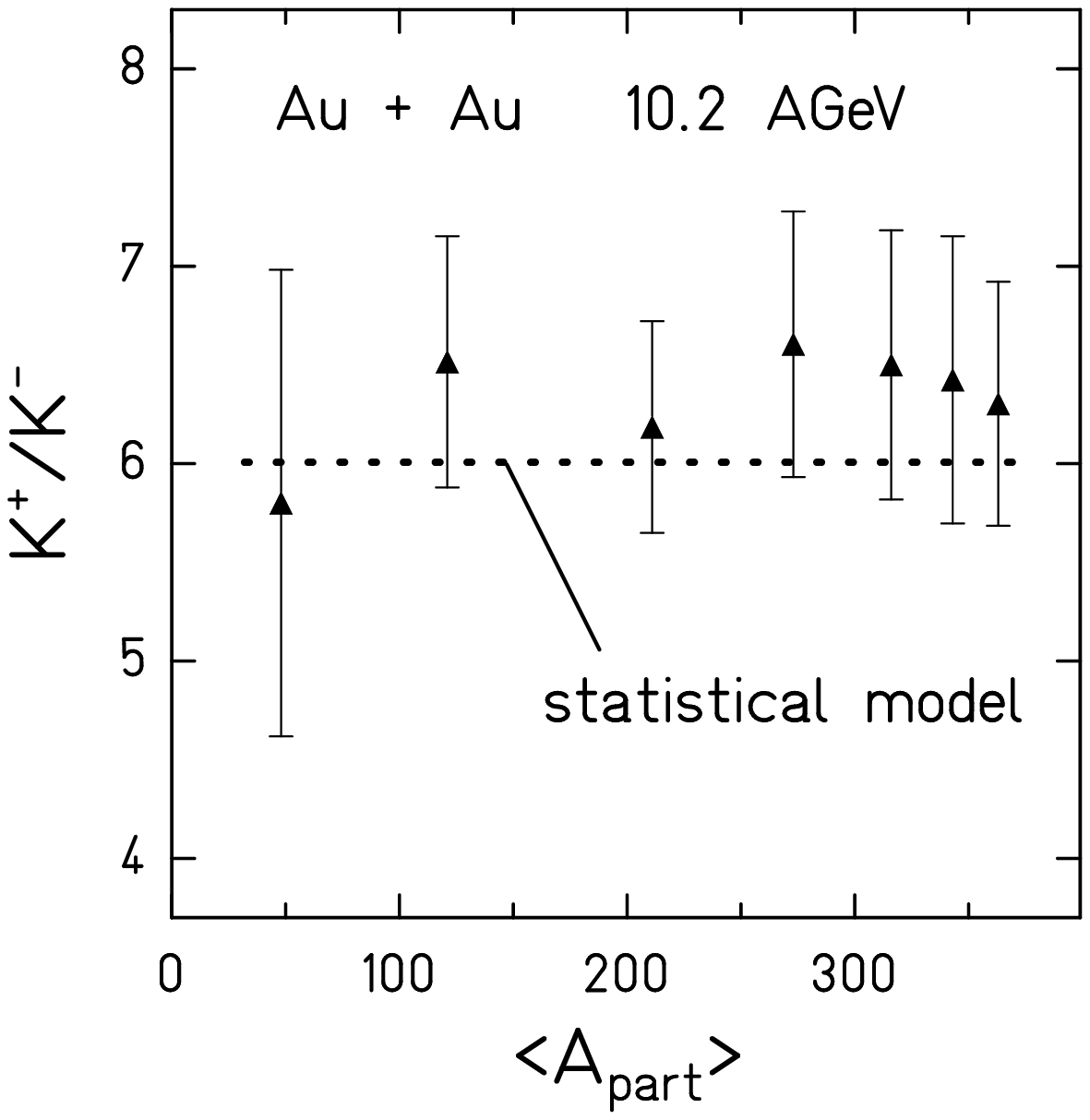}}
\mbox{\epsfig{width=7.2cm,file=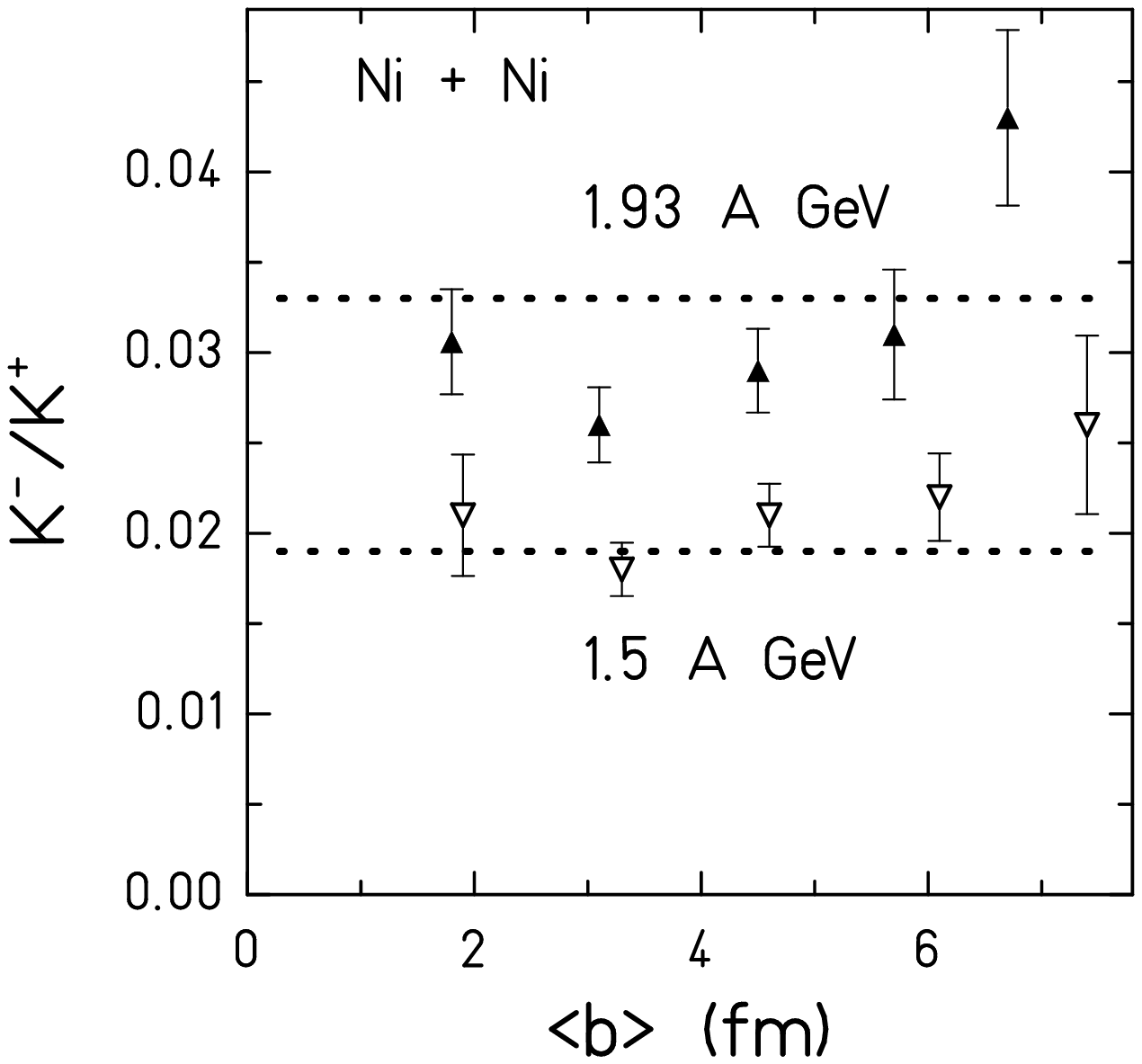}}
\caption{Ratio of $K^+$ to $K^-$  as a function
of the number of participants $A_{part}$ 
(left for Au+Au at 10.2 $A\cdot$GeV~\cite{Ahle99}, 
right for Ni+Ni at 1.5 and 1.93 $A\cdot$GeV~\cite{Marc}
together with the statistical model (dashed line) evidencing the
independence of $<A_{part}>$.}
\label{KPKMapart}
\end{figure}

Furthermore, 
the two formulae predict that the $K^+/K^-$ ratio for a given collision
should not vary with centrality as $V$ cancels in the ratio.
Indeed, this has been observed in Au+Au collisions between 4 and 
10.2 $A\cdot$GeV~\cite{Ahle, Dunlop} and also at SIS energies~\cite{Marc}.
Two examples are shown in Fig.~\ref{KPKMapart}.
This independence of centrality is rather astonishing 
as one expects an influence
of the different thresholds and the density variation with centrality,
especially at 1.93 $A\cdot$GeV where the $K^+$ production is above,
the $K^-$ production below their respective thresholds.

Transport-model calculations and other estimates show clearly that 
strangeness equilibration requires a time interval of 
40 -- 80 fm/$c$~\cite{Koch86, Brat00}.
While the statistical model is quite successful describing the
particle yields, including strange particles,  this cannot be taken as proof
for chemical equlibration. Especially in case of the $K^+$ production,
no strong absorptive channel seems to be available. This point clearly deserves
further attention. 
At these low incident energies the strange quarks are found only in
a few hadrons. The $\bar s$ quark is found only in $K^+$, while
the $s$ quark will be shared between $K^-$ and $\Lambda$ (or other hyperons).
The latter sharing might be in chemical equlibrium as 
the reactions
$$\pi^0 + \Lambda \rightleftharpoons p + K^- \quad \rm{or} \quad
\pi^- + \Lambda \rightleftharpoons n + K^-$$
 are strong and have only 
slightly negative Q-values of -176 MeV.
If these reactions are the dominating channel, the law of mass action
might be applied giving for the respective concentrations
\begin{equation}
\frac{[\pi] \cdot [\Lambda]}{[K^-] \cdot N} \, = \, \kappa .
\end{equation}
As the number of $K^-$ relative to $\Lambda$ is small, $[\Lambda]$ can be
approximated by $[K^+]$ and rewriting gives
\begin{equation}
\frac{[K^-]}{[K^+]} \propto M(\pi^0 + \pi^-)/A.
\end{equation}

\begin{figure}[h]
\begin{minipage}[t]{5.1cm}
\mbox{\epsfig{width=7.6cm,file=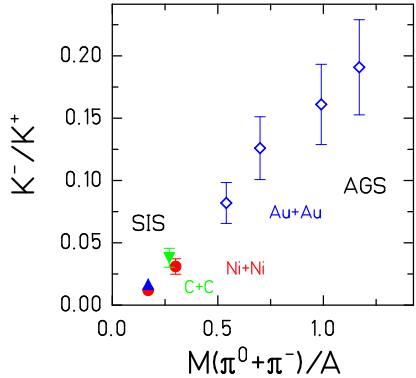}}
\end{minipage}
\hfill
\begin{minipage}[b]{5.9cm}
\vspace*{-3cm}
\caption{The $K^-/K^+$ ratio as a function of the
pion multiplicity $M(\pi^-+\pi^0)/A$ as a test of the law of mass action.
Preliminary data.}
\label{Massaction}
\end{minipage}
\end{figure}

Figure~\ref{Massaction} tests the validity of this relation 
at SIS and AGS energies showing a rather linear rise 
of $K^-/K^+$ with $M(\pi^-+\pi^0)/A$ reflecting a constant value of $\kappa$.
 
\section{Beyond the Statistical Model}

While the presented results exhibit an astonishing agreement with
the statistical model there are clear experimental observations which 
are beyond the statistical description indicating new phenomena.
These results are presented by the following talks:
\begin{itemize}
\item The emission of positively charged kaon, measured by the KaoS
Collaboration,
exhibits an elliptic flow
which is hard to understand as the $K^+$ interaction with nuclear matter 
is small~\cite{Shin}. 
Only by including a repulsive $K^+N$ potential
the experimental data can be described. Talk by {\bf Y. Shin}.
\item The measured $p_t$-dependence of the directed flow
of $K^+$ in Ru+Ru and Ni+Ni collisions near threshold,
studied by the FOPI collaboration, 
can only be decribed
when using a repulsive $K^+N$ potential~\cite{Crochet}.
Talk by {\bf P.~Crochet}. An extremely strong antiflow signal of $K^0$
is found in Au+Au collisions at 6 $A\cdot$GeV by the 
EOS collaboration~\cite{Chung}. 
Talk by {\bf P.~Chung}. 
\end{itemize}

Direct experimental evidence for the time evolution
of pion emission is presented based on the shadowing of spectator
matter in certain space-time regions~\cite{AW99}.
In peripheral collisions of
Au+Au at $1.0~A{\cdot}$GeV incident energy
the moving spectator matter acts like a shutter of a camera shielding the pions,
i.e.~modifying the pion emission pattern according to the spatial distribution
of the spectator matter at the time of the pion freeze out.
The motion of the spectator serves as a calibrated clock.
This analysis suggests that in Au + Au collisions at 1~$A\cdot$GeV
most of the high-energy pions freeze out within
10 fm/$c$ after the first touch of the nuclei.
Low-energy pions do freeze out later and 
over a longer time interval~\cite{AW99}.

\section{Summary}

The measured multiplicities of $K^+$ and $K^-$ production at subthreshold 
energies rises more than linearly with system mass $A$ and with the number of 
participating nucleons $A_{part}$.
The spectra do not show deviations from Boltzmann shapes.

A global survey of the spectral shapes and yields
of the emitted pions, $K^+$ and $K^-$
points towards an interpretation within a statistical concept.
This model takes into account exact strangeness conservation, i.e.~the
associate production of two strange particles (e.g.~$K^+$ and $\Lambda$).
It does not only explain the various particle ratios (except $\eta$), but also
describes the very different $A_{part}$-dependences of the $K^+$ and $\pi$
multiplicities.

Despite  the apparent success of the statistical model of
particle production under the assumption of thermal and
chemical equilibration and using masses of free particles,
the present understanding of hadronic
interactions contradicts chemical equilibrium
for strange particles~\cite{Brat00}.
While for $K^-$ the channel
$K^- + N \rightleftharpoons \Lambda + \pi$ is open, for $K^+$
no corresponding channel is available needed to understand how
equilibrium can be obtained.
The kaon yields in the statistical model are based
on free masses, while transport model largely under-predict
the measured $K^-$ yields when using bare masses. Only if in-medium
masses for $K^-$ are used, which are considerably lower, agreement with
the measured rates can be achieved.
This discrepancy seem
 to put into question our present understanding
of interactions at the high densities reached in heavy ion collisions.

Furthermore, the observation of a strong elliptic flow of $K^+$,
and of pronounced $p_t$-dependence of a directed flow can only be described
using a repulsive $K^+N$ potential.
                                 
The discrepancy between the two controversial interpretation
needed to describe the data might  disappear when
the collision process
is understood as a dynamical evolution; 
in some observables, e.g.~particle ratios,
the dynamics is lost. However, when looking at small areas of the phase space
the influence of a dynamical evolution could be traced back.\\

\end{document}